# Distinct changes of genomic biases in nucleotide substitution at the time of mammalian radiation

*Peter F. Arndt[1,2,*], Dmitri A. Petrov[3], and Terence Hwa[1]*

[1]*Physics Department and Center for Theoretical Biological Physics, University of California at San Diego, La Jolla, CA 92093*

[2]*Institute for Theoretical Physics, Cologne University, 50937 Cologne, Germany*

[3]*Department of Biological Sciences, Stanford University, Stanford, CA 94305*

[*]Corresponding author: email: peter.arndt@uni−koeln.de

Phone:  +49 221 470−5203

Fax:     +49 221 470−5159

Keywords: nucleotide substitution, CpG−methylation, repetitive elements, GC−content

Running head: Changes of biases in nucleotide substitution

Abbreviations:   RE − repetitive element

MYA − million years ago

ML − maximum likelihood

**Abstract**

Differences in the regional substitution patterns in the human genome created patterns of large−scale variation of base composition known as genomic isochores. To gain insight into the origin of the genomic isochores we develop a maximum likelihood approach to determine the history of substitution patterns in the human genome. This approach utilizes the vast amount of repetitive sequence deposited in the human genome over the past ~250 MYR. Using this approach we estimate the frequencies of seven types of substitutions: the four transversions, two transitions, and the methyl−assisted transition of cytosine in CpG. Comparing substitutional patterns in repetitive elements of various ages, we reconstruct the history of the base−substitutional process in the different isochores for the past 250 Myr. At around 90 Myr ago (around the time of the mammalian radiation), we find an abrupt 4− to 8−fold increase of the cytosine transition rate in CpG pairs compared to that of the reptilian ancestor. Further analysis of nucleotide substitutions in regions with different GC−content reveals concurrent changes in the substitutional patterns. While the substitutional pattern was dependent on the regional GC−content in such ways that it preserved the regional GC−content before the mammalian radiation, it lost this dependence afterwards. The substitutional pattern changed from an isochore−preserving to an isochore−degrading one. We conclude that isochores have been established before the radiation of the eutherian mammals and have been subject to the process of homogenization since then.

Different regions of the human genome show large variation in GC−content from 30% to 60% at scales exceeding hundreds of kilobases (Filipski, Thiery, and Bernardi 1973; Bernardi 2000; Eyre−Walker and Hurst 2001). The origin, timing and implications of this so called "human isochore structure" is still controversial, partly due to a number of technical issues that have made it difficult to reconstruct the history of nucleotide substitutions. The difficulties in assessing substitutional biases in vertebrate genomes that acted on coding and non−coding sequences far in the past come from two sources. One is that many substitutional processes in non−coding regions are fairly fast, leading to multiple changes at the same site (so−called "multiple hits"), which quickly obscure the true pattern of substitution. The other is the strong *neighbor−dependence* of some substitution processes, especially the methylation−assisted transition of cytosine found in 5'−CG−3' (CpG) pairs, leading to a sharply elevated (by as much as 10−fold) rate of substitution from CpG to TpG and CpA (Coulondre et al. 1978). Such neighbor−dependence exacerbates the problem of multiple hits and severely complicates any inference of substitutions beyond the very immediate past (i.e., more than 10−15 Myr ago).

To address these issues we developed a maximum likelihood (ML) method that simultaneously estimates the substitution rates corrected for multiple hits, for the four transversions, two transitions, and the CpG−based transition process. This method is based on a recently developed treatment of neighbor−dependent substitutions (Arndt, Burge, and Hwa 2002), which is an extension of Kimura's approach along with its generalizations (Kimura 1981; Hasegawa, Kishino, and Yano 1985) to include the CpG−based transition process. Importantly, it is capable of recovering the true patterns of substitution going far back in time, given the knowledge of the ancestral sequence and a sufficient amount of data.

In the case of the human genome, sufficient amount of data is indeed available in the form of exceptionally prolific interspersed repetitive elements (REs), which have been inserted into the human genome in bursts at various stages during evolution (Britten et al. 1988; Jurka and Smith 1988; Britten 1994; Kapitonov and Jurka 1996; Mighell, Markham, and Robinson 1997). Every burst generated numerous copies of an ancestral "Master sequence", which can be reconstructed easily from the available data (Jurka 2000). The Master sequences was not the same for every burst; this way various sub–families of RE have been generated at different times in evolution. Most of the REs reside in the intergenic regions and are believed to be functionally neutral.

The oldest families of REs that can be identified in the human genome have been inserted over 200 Myr ago (MYA). Naturally, older elements accumulated more base substitutions than younger ones. Therefore, a careful comparison of the frequencies of substitutions observed in REs of different ages can tell us the history of the substitutional processes extending far into the past. This analysis applied to all REs found in the human genome provides us with the information about the genome– wide averaged background substitutional processes at different times. Using the same analysis and applying it to REs found in specific regions we could reconstruct the region–specific substitutional processes, as well. We were able to reveal different substitution patterns in regions with different background GC–content. These patterns changed in time and therefore provide us additional information on the origin and timing of the human isochore structure.

## Materials and Methods

**Data Collection.** This work is based on the analysis of ~2,800 Mbp of autosomal sequence data from the human genome, available at GenBank (Build 28; February 11, 2002). The "RepeatMasker" (http://repeatmasker.genome.washington.edu) was used to identify copies of various REs and to classify them into families and sub–families as they appear in the RepBase database (Jurka 2000). With this procedure, 46% of the genome was identified to be either a part of an RE or a sequence of particularly low complexity. The subsequent analysis was focused on elements from the 42 subfamilies of retrotransposons (see Table 1), including 7 sub–families of Alu's (Jurka and Milosavljevic 1991), 32 subfamilies of the LINE element L1 (Smit and Riggs 1995), as well as older elements such as L2, L3 (Jurka 2000), and MIR (Smit et al. 1995). We excluded partial alignments shorter than 250 bp for the Alu's, or shorter than 100 bp for the others. The number of remaining alignments and their average lengths are listed in Table 1 and sums to a total of 410Mbp of sequence data (> 10% of the human genome) that was used in the analysis. We further verified that the identification of the oldest REs (MIR, L2, L3) is well within the capability of RepeatMasker. We synthetically aged the ancestor sequences of these RE families (see below) to a significantly higher degree of divergence (including indels) than what was found in the data and verified that we could identify these using the RepeatMasker. To avoid potential biases on the set of identified repeats by the regional GC−content, the "−gc 43" option of the RepeatMasker was turned on, and the scoring system appropriate for the genome−wide average GC−content of 43% was chosen. Note, however, that whether the "−gc 43" option was on or not did not significantly affect the results. For the output of local gapped alignments of each identified copy with its ancestor sequence needed for our further analysis we turned the "−a" option on as well.

Due to the large number of copies of REs in each subfamily, it was also possible to subdivide these copies into seven groups according to the GC−content of regions flanking each identified copy. In counting the GC−content of the flanking regions, we masked away the sequences of other REs appearing in the regions. The modulation of the length of the flanking regions from 2kb to 10kb did not affect the results. We verified that the shift in GC−content in the last 100 Myr due to the current substitution pattern is less than *5%* (results not shown). Thus, the regional GC−content as measured in today's genome reflects approximately those at about 100 MYA. For the genome−wide analysis of substitution frequencies, we further accounted for the different densities of REs in the different isochores (Gu et al. 2000) and randomly discarded identified copies of REs such that the total amount of repetitive sequence was proportional to the total amount of genomic sequence of a given GC−content.

**Data Analysis.** In the analysis, the "star" phylogeny was assumed for each subfamily of REs. This assumption is based on the known biology of retrotransposons, and supported by previous phylogenetic analysis of the transposons (Britten et al. 1988; Jurka and Smith 1988; Jurka and Milosavljevic 1991). The assumption is invalidated if a significant fraction of the REs is generated by duplication. However, the latter is estimated to affect under 10% of the REs (J. Jurka, personal communication), making the star phylogeny a reasonable starting point. Given the degree of sequence divergence of the copies of REs (Table 1) we assume that each copy evolves as a unique sequence. Further, the extremely low divergence of the youngest REs (Jurka 2000) in the human genome suggests that retrotranscription errors can be safely ignored. The ancestor sequence for each subfamily is taken from the RepBase (Jurka 2000) and verified by a ML−based reconstruction method.

**Estimation of substitution frequencies.** We used a maximum likelihood (ML) approach to estimate the substitution frequencies for each sub−family of REs. The observed data is given by the pair−wise gapped alignments of each identified copy of a RE in the genome ($\beta_1$,?, $\beta_M$) with its ancestral sequence ($\alpha_1$,?, $\alpha_M$) from the RepBase (Jurka 2000). Here Greek letters represent the nucleotides: A, C, G and T, and $M$ is the length of the alignment. From the alignments of a particular subfamily of RE, we count the number of observed substitutions $\alpha_i \rightarrow \beta_i$, recording also the neighboring bases in the ancestor, $\alpha_{i-1}$ to the left and $\alpha_{i+1}$ to the right. We disregard a site $i$ if any of the $\alpha_{i-1}$, $\alpha_i$, $\alpha_{i+1}$, $\beta_i$ is an insertion or deletion in the alignment. We denote these counts by $N(\beta|\alpha_L, \alpha, \alpha_R)$ and the set of all counts for a subfamily by $\{N\}$.

To implement the maximum likelihood approach, we need to specify a substitution model describing the observed data. We chose a general model comprising all possible single nucleotide transversions (8) and transitions (4) as well as the CpG−based transition, CpG$\rightarrow$CpA/TpG. Each process and its complementary process are assumed to occur with the same substitution frequency per site. Hence the substitution model is parameterized by a set of seven frequencies, collectively denoted as $\{\omega\}$. The likelihood of observing the data $\{N\}$ under a given model with parameters $\{\omega\}$ can be approximated by where $P(* \beta * |\alpha_L, \alpha, \alpha_R; \{\omega\})$ is the probability that a sequences of three bases set up initially with the configuration $\alpha_L$, $\alpha$, $\alpha_R$ evolves under the model $\{\omega\}$ to a state with the base $\alpha$ substituted by $\beta$. To include effects due to multiple– and back–substitutions of the same base, this probability is calculated by iterating a set of 64 differential equations that encodes the time evolution of the three nucleotides taking the set $\{\omega\}$ as parameters (Arndt, Burge, and Hwa 2002). The likelihood $L(\{N\}|\{\omega\})$ can then be maximized using standard algorithms (Press et al. 1992), adjusting $\{\omega\}$ to find the set of substitution frequencies which most likely describes the observed data. The typical error of the ML method is estimated by bootstrap (Press et al. 1992). Due to the large amounts of sequence data for every family of REs, the estimated errors are small. In the following, error bars smaller than the symbol size in the figures presented  will be omitted.

The heart of our method is the recently developed *neighbor–dependent* substitution model (Arndt, Burge, and Hwa 2002). This method is an extension of Kimura's approach and its generalizations (Kimura 1981; Hasegawa, Kishino, and Yano 1985) to include the CpG–based transition process. It was assumed that the REs were selectively neutral, so that the two DNA strands can be treated in the same way. The model comprised the 4 transversions, 2 single–nucleotide and the CpG–based transitions, as well as all the secondary processes involving multiple– and back–substitutions. A special case of this model in which all the transition rates were equal and all the transversion rates were equal was solved analytically (Arndt, Burge, and Hwa 2002). For the more general case at hand involving all 7 substitution frequencies, the principle of the likelihood calculation was the same as described by Arndt et al (2002), but the calculation became more tedious and was performed with the help of a computer. The corresponding program is available upon request. Undoubtedly, the inclusion of the neighbor–dependent substitution process made the analysis more complicated. However, the results obtained were more sensitive and reliable at longer evolutionary time scales, especially when compared to the alternative method of "direct counting" of CpG $\rightarrow$ CpA/TpG events. The performance of this approach  has been evaluated quantitatively as will be shortly described below.

**Reconstruction of the ancestral sequence.** In our analysis, we assumed that the RE sequence appearing in the RepBase was the correct ancestral sequence (Jurka 2000). We tested this assumption by reconstructing the ancestral sequence from all copies found in the human genome, using the ML approach with the above substitution model. In contrast to the above estimation of substitution frequencies, here we want to find the most likely ancestral sequence ($\alpha_1, ?, \alpha_M$), given the observed copies and our substitution model. The likelihood to be maximized is again $L(\{N\}|\{\omega\})$ but now one has to adjust the ancestral sequence $\alpha_i$ (which changes the numbers $\{N\}$) while keeping the substitution frequencies $\{\omega\}$ fixed. Initially the frequencies $\{\omega\}$ are unknown. We use the naïve consensus sequence and the ML analysis described above to get an estimate for these frequencies. Subsequently, we construct the ML−ancestor by varying the ancestor sequence and maximizing $L(\{N\}|\{\omega\})$ using the $\{\omega\}$ just determined. We then use the new ancestor sequence to get a better estimate of the frequencies. After three iterations, this procedure converges to a unique ML−ancestor sequence and estimated substitution frequencies. We tested this method by independently evolving many copies (in number comparable to the number of copies of REs found in the human genome) of a synthetic ancestral sequence. We were able to reconstruct  this ancestor sequence exactly, even if all CpG disappeared from the naïve consensus of the evolved sequences due to  the high CpG−based transition frequency. Comparing the ML−ancestor and the ancestor in the RepBase, we observe differences for less than 1% of the sites.

**Performance Evaluation.** To test the performance of the ML method, we synthetically aged sequences with known substitution frequencies starting from an ancestral sequence. We used the following stochastic evolution procedure: Pick a base at random and allow one of the seven substitution processes we consider to occur with probability that is 1/1000 of the corresponding substitution frequency $\omega$; repeat until every base was visited 1000 times on average. Scaling all $\omega$'s by a single factor while keeping their ratios fixed, we can generate sequences evolved for various amounts of times with different degrees of divergence from the ancestor. The ratios themselves are the relative substitution frequencies and are independent of the amount of time the sequences have been aged for.

To test the accuracy and limitation of our ML method in estimating the substitution frequencies, we synthetically evolved 20,000 copies of the ancestral MIR sequence (5 Mbp in total, about 1/10 of the amount of MIR sequences found in the human genome) for varying periods of time $t$, with equal transversion rates and differing transition rates (A:T$\rightarrow$G:C rate=3x, G:C$\rightarrow$A:T rate=5x, and CpG− based transition rate=40x that of the transversion rate respectively). These rates are chosen to be similar to those found in our later analysis. We perform the ML analysis for sequences at each age $t$, and plot the 3 transition frequencies (solid symbols) obtained against the average of the transversion frequencies for each age in Fig. 1. We observe fixed ratios between the transition and transversion frequencies, with the slopes reflecting very accurately (within 1% relative errors) the relative substitution rates used in aging. Estimates of the same frequencies using the direct counting method are indicated in Fig. 1 by the open symbols. The result is especially poor for the CpG−based transition frequencies (the open diamonds): For sequences with transversion frequency above 0.02, this DC−estimated CpG transition rate saturates to 0.5, indicating that most CpG's had decayed into either CpA's or TpG's.

In our ML analysis, we assume that the relative substitution rates are time independent. However, we want to demonstrate that even if the substitution rates are time dependent, the ML analysis can still be used to give fairly accurate results, primarily due to the very large amount of sequence information in our dataset. Again we start with the 20,000 copies of the MIR ancestor sequences, and age them for various periods of time $t$. But this time, we use two different sets of substitution rates, referred to as the "present" and "past" rates with respect to a time $t_0$ when the rates are changed from one to the other. For $t>t_0$, we age the sequences with the past rates for a period $t−t_0$, and then age them with the present rates for a period $t_0$. For $t<t_0$, we age with the present rates for the entire period $t$. We then apply our ML method to analyze sequences aged for each time period $t$, estimating substitution frequencies with the implicit assumption that the rates are time independent.

We first consider the effect of a 5−fold increase in the CpG−based transition rate (from 8x to 40x the transversion rate), while keeping the relative single−nucleotide transition rates the same as before. The change is applied at a time $t_0$ corresponding to an average transversion frequency of *0.025,* to mimic an observed effect to be described below. The frequencies estimated by ML for various aging time are shown in Fig. 2. Here and below, it will be convenient to use the average transversion frequency as the unit of aging time $t$. For $t<t_0=0.025$, the sequences don't know about the rate change at t0, and the estimated substitution frequencies are the same as those obtained before (as shown in Fig. 1). For $t>t_0$, the estimated CpG−based transition frequencies (the diamonds in Fig. 2) become very different, while the other two transition frequencies (the triangles) follow the same trend as for $t<t_0$. When plotted against the average transversion frequencies, the CpG−based transition frequencies exhibit a sharp change at $t=0.025$ and fall reasonably well onto two straight lines. We see that the slopes for the data points in the two regions reflect the two relative substitution rates used in the aging process (slopes of the solid lines in Fig. 2). Even for the less reliable "past" rate, we obtain a relative error of $< 10\%$. Of course, the actual error size depends on the total amount of sequence data for each time $t$.

A similar test was performed to check whether a change in the single−nucleotide transition rates could be captured using the same type of analysis. In Fig. 3, we present the results of this test subjecting 20,000 copies of the ancestral MIR sequences to substitution processes observed in later analysis. We find that the ML analysis gives excellent estimates for the relative substitution rates both before and after the time $t_0$. In what follows, we will interpret the different slopes in different regions of the transition−transversion plots as estimates of the different relative transition frequencies at different times. Of course we could have extended our ML analysis to include time−dependent rates from the beginning. However, this process would have required adjustment of many more parameters and become very time consuming, while the approach we propose is simple to implement and sufficiently accurate for the large amount of data we have.

## Results

**Genome−wide averaged substitution pattern over the past 250 MYR.** To investigate the *history* of substitution processes during the course of human evolution, we extracted from the human genome numerous copies of the commonly encountered families of retrotransposons, the SINEs (Alu, MIR) and LINEs (L1, L2, L3), see Table 1 for a complete listing. For each of the 42 subfamilies of REs obtained, we computed the average number of substitution events per site (corrected for back and multiple substitutions) for each of the 3 transitions (the 2 single−nucleotide transitions and the CpG−based transition CpG → CpA/TpG) and 4 transversion processes using our maximum likelihood method; the results are referred to as transition and transversion "frequencies" respectively. In Fig. 4, we plot the 3 transition frequencies against the average of the four (very similar) transversion frequencies for each subfamily. From this figure, we observe first that the two single−nucleotide transition frequencies (the triangles) show remarkably *linear* dependence on the average transversion frequencies. This suggests that the genome−wide averaged single−nucleotide substitution pattern (as defined in *Methods*) has not changed since the time the oldest elements (L3, L2, MIR) entered the genome. Fitting these two transition frequencies to straight lines, and identifying the slopes as the transition rates (relative to the average transversion rate), we find that the A:T → G:C transition (up−pointing triangle in Fig. 4) occurs *2.74 ± 0.04* times more frequently, and the G:C → A:T transition (down−pointing triangle) *5.5 ± 0.1* times more frequently than an average transversion event. Given the linearity of the data, it is tempting to assume that the single−nucleotide substitution *rates* are time−independent throughout the time span studied, and use the horizontal axis of Fig. 4 as the "time" axis. Calibrating the time scale using estimates of the absolute insertion time of the different Alu sub−families (Kapitonov and Jurka 1996), we find each unit of *t=0.01* in the average transversion frequency to correspond to approximately *35* Myr, with the entire dataset spanning nearly *250* Myr.

**Change in CpG−based transition in the ancestor of eutherian mammals.** The corrected frequencies obtained for the CpG−based transition process (diamonds in Fig. 4) present a big surprise: The data are almost certainly ($P = 10^{-13}$) not described by a single straight line. Instead, there are two regimes characterized by very different slopes. This finding is supported by analysing REs across different families of SINEs and LINEs, including the Alu's, which insert preferentially in GC−rich regions as well as the L1's, which prefer AT−rich regions. We verified that this finding is not an artefact of the ML analysis: extensive simulation (see Methods) shows that our method is capable of detecting transition frequency for the CpG−based process up to 4−5 times that of the highest frequencies found in our data. To quantify the extent to which the transition rates changed in the past, we divided the data into the two sets of young and old REs with respect to a threshold value $t_0$. Subsequently $t_0$ was adjusted such that the sum of the squared residuals of linear regressions to the data in both sets is minimal. For the data on the CpG–transition process (as shown by the diamonds in Fig. 4), this minimum was found for $t_0=0.025$, with a slope of *39.5 ± 2.6* for the young elements and *8.4 ± 2.5* for the old elements (relative to the average transversion rate). As demonstrated by extensive simulations described in Methods, we can identify the two slopes with the relative rates before and after $t_0$. This leads to the conclusion that a 4– to 8–fold increase in the CpG–based transition rate occurred at $t_0=0.025$. On the other hand, we could not detect any significant change in the rates between the young and old elements for the two single–nucleotide transition processes (triangles in Fig. 4). The single straight lines fit the rates of both transitions relative to the average transversion rate remarkably well (*R = 0.99*), suggesting that a split of the data into sets of young and old elements is not justified.

Intriguingly, $t_0=0.025$ (~ 90 MYA) corresponds roughly to the time of the mammalian radiation (Hedges et al. 1996; Kumar and Hedges 1998; Easteal 1999; Murphy et al. 2001). Corroborating this inference is an independent observation in Fig. 4 that the CpG–based transition frequencies of nearly all the L1Pxx elements (solid square symbols, corresponding to the L1's found in all primates) fall on the steep segment, while the CpG–based frequencies of all the L1Mxx elements (solid diamond symbols, corresponding to L1's found in all eutherian mammals) (Smit and Riggs 1995) fall on the flatter segment. One possible mechanism for the abrupt increase of the CpG–based transition rate (but not the neighbor–independent C $\rightarrow$ T transition rate) is a hypothetical elevation of methylation activity in germ–line cells at the time of mammalian radiation.

**The history of substitution in different isochores.** To gain some insight into the strong regional GC–content variations of the human genome, we repeated the above analysis separately for REs found in regions with different base composition. Specifically, we partitioned the REs according to the base composition of the region flanking individual copies into 7 equally–sized bins of GC–content ranging from 30% to 60%. A plot similar to Fig. 4 was generated for each of the GC–content bins and analyzed separately. As expected, we find the abrupt upward shift in the rate of CpG–based transition for every bin of GC−content, occurring at approximately 90 MYA (data not shown). More surprisingly, the rates of single−nucleotide transitions also showed time dependence. As shown in Fig. 5, both transition frequencies are nearly independent of GC−content for transversion frequencies *<0.02*, i.e., in the recent past, while clear dependence on GC−content can be seen for the two transitions at much earlier times, e.g., for transversion frequencies *>0.03.* Moreover, the dependences of GC−content in the distant past are opposite for the two transitions: The rates of GC−enriching transitions were higher in the regions of higher GC−content (Fig. 5a) while the opposite was the case for the AT−enriching transitions (Fig. 5b). Similar results have already been found by Lander et al (2001). It thus appears that the GC−dependence of the substitution pattern also underwent a change. Intriguingly, this change coincides in time (~90 MYA) with the upward shift of the CpG−based transition rate and the mammalian radiation. Note that this change is not a trivial consequence of the shift in the genome−wide CpG−based transition rate shown in Fig. 4, since the different substitution processes are treated separately in our ML analysis.

We quantified all of the substitution rates in the recent and distant past by fitting the data such as those in Fig. 5 to straight lines, separately for those with transversion frequencies below 0.02 and above 0.03. This was done for all 7 bins of GC−content, allowing us to analyze the dependences systematically. In the distant past, i.e., prior to ~90 MYA (Fig. 6), the GC dependence of substitution rates appears to have been preserving or generating both the GC−rich and GC−poor isochore structures: AT−enriching substitutions (down−pointing triangles) were predominant in the GC−poor isochores, and the reverse was true for the GC−rich isochores (Fig. 6a). We examined the degree of preservation quantitatively by computing the stationary GC−content corresponding to each of the substitution patterns for each bins of GC−content in Fig. 6a. This computation is performed according to the method described in (Arndt, Burge, and Hwa 2002). The results obtained (Fig. 6b) support the conclusion that back at ~90 MYA, steady state had been reached and the GC−content of each region was indeed preserved by the regional substitution patterns. Note also that the very good agreement between the observed GC−content of the flanking regions and the predicted stationary GC−content from the patterns of substitutions within the REs validates two important aspects of our approach. First, it shows that we can faithfully estimate patterns of substitutions within REs going back $90 - 250$ Myr. It also shows that the patterns of substitution within REs are fair estimates of substitutional processes in their flanking regions.

The situation has been very different in the recent past, i.e. since ~90 MYA. As shown in Fig. 7a, the substitution patterns obtained are very different from those in Fig. 6a ($p<10^{-6}$). For the recent past, there is only weak dependence of the substitution frequencies on the GC−content, a result which has also been reported recently by Yi et al. (2002). However, despite these dependencies, and especially since the frequencies of AT−enriching substitutions (down−pointing triangles in Fig. 7a) always exceed those of the GC−enriching substitutions (up−pointing triangles), it is clearly not possible to preserve the regional GC−content with such a substitution pattern. The plot of stationary versus regional GC−content (Fig. 7b) is computed using the rates given in Fig. 7a. The most striking feature of this plot is the inference that the high−GC isochores are currently disappearing from the human genome. Assuming the persistence of the present substitution pattern, the GC−content will be homogenized in the range of $35 - 40\%$ in about 400 Myr (calculation not shown). The ongoing loss of high GC isochores is also implicit in the results derived from the analysis of transposable elements by Lander et al. (2001). Similar conclusions have been recently noted by Duret et al. (2002) who estimated the rates of synonymous substitutions in orthologous genes in closely related species within primates, certiodactyls and rodents. The parallel shift toward lower GC−contents in several orders of eutherian mammals suggests that the shift occurred in their common ancestor (Duret et al. 2002). Our analysis agrees with this interpretation and further quantifies it by showing that the shift occurred around the time of mammalian radiation.

## Discussion

The ability to directly estimate the rates of all the major types of substitutions so far back in time allows us to shed light on several contentious issues in the evolutionary history of the human genome.

The inferred substitution patterns provide strong support for the inference by Galtier and Mouchiroud that the ancestor of eutherian mammals had a very pronounced variation in GC content, more similar to that found in humans than in rodents (Galtier and Mouchiroud 1998). Our results also suggest that the substitution pattern leading to the eventual establishment of high–GC isochores evolved in the common ancestor of amniotes, resulting in the parallel evolution of the strongly heterogeneous GC–content in mammals, birds and reptiles (Bernardi 1990; Bernardi, Hughes, and Mouchiroud 1997; Hughes, Zelus, and Mouchiroud 1999). Thus it is very unlikely that the high GC–content isochores evolved as an adaptation to homeothermy in either mammals or birds (Bernardi 1993). Our study also shows that the currently fast rate of CpG transitions in the mammalian lineages is unlikely to be an important cause in the generation of isochores – CpG transition rate was much lower at the time isochores were generated and it increased at the time when isochores started to degrade.

Our estimates of the recent pattern of substitution agree with those derived by other authors by studying evolution at synonymous sites in pairs of mammalian species (Duret et al. 2002; Rosenberg, Subramanian, and Kumar 2003) and by studying SNPs in the human populations (Eyre–Walker and Hurst 2001; Duret et al. 2002; Webster, Smith, and Ellegren 2003). In any case, the bigger estimation error is expected in the older, pre–90 MYA patterns of substitution. However, those are the patterns that produce a very good fit between the observed and the expected GC content in the human genome. Therefore, we believe that our main results are robust to the possible errors in the estimation procedure. It is noteworthy that although the GC content expected based on pre–90 MYA patterns fits well the observed GC content, it consistently underestimates it by a small amount (Fig. 6b). This suggests the existence of another GC–enriching process acting fairly uniformly on most sequences in the genome. The possible candidates include insertions of GC–rich sequences (pseudogenes or REs) and an AT–bias in small deletions or a GC–bias in small insertions that we have not taken into account in our procedure.

If the high–GC isochores have been continuously degraded over the last 90 MYR, why do we still see the pronounced isochore structure in the human genome? We believe this is primarily because the degrading process is very slow. Based on the estimated substitution rates, we expect that high–GC isochores had at most 5% higher GC–content 90 MYA. This would imply that the faster–evolving mammalian lineages should have the more muted isochore pattern – the expectation is borne out in the mouse lineage (Mouchiroud, Gautier, and Bernardi 1988).

How can we explain the sharp reduction in the GC–dependence of *all* transition and transversion rates (Fig. 5 and 7), without any apparent change in the genome–wide rate of transitions and transversions at approximately 90 MYA (Fig. 4)? One plausible scenario involves a combination of the regional variation in substitution rates [driven by either mutational biases (Sueoka 1988; Wolfe, Sharp, and Li 1989), or consistent differences in the strength of selection (Bernardi 1986; Eyre–Walker 1999) or frequency of biased gene conversion (Holmquist 1992; Eyre–Walker 1993)] and increased chromosomal rearrangement activities in the mammalian lineage. Imagine that some regions in the genome (for instance centromeric and telomeric regions) have consistently higher rates of AT to GC substitutions. Without chromosomal rearrangements, the DNA sequences in these regions will eventually reach steady state and attain a higher GC content consistent with the local substitution rates. Suppose the rate of chromosomal rearrangements increased significantly after steady state was reached, e.g., at around 90 MYA. Then until a new steady state can be reached (a process which would take several hundred Myr), some RE's in GC–rich segments would be subject to low AT $\rightarrow$ GC substitution rates while others would continue to have high AT $\rightarrow$ GC rates, and similarly for RE's in GC–poor segments. This would remove the apparent GC–dependence of the substitution rates since 90 MYA, and explain why all transition and transversion rates changed their GC–dependences synchronously at that time while the genome–averaged substitution rates hardly changed at all. An alternative scenario is the occurrence of an isolated episode of increased chromosomal rearrangements at around 90 MYA. This would also produce the observed shift in the GC–dependence of the single–nucleotide substitution patterns. However, under this scenario, we would expect the isochore structure existing prior to 90 MYA to re–establish itself rather than becoming completely degraded after several hundred Myr.

This model requires that the rate of chromosomal rearrangements in the amniote ancestor be lower than what it is in eutherian mammals. It predicts that the mammalian lineages with higher rates of chromosomal rearrangements should show more muted isochore structure, with less variation of the GC content across the genome. The mouse lineage does appear to have a higher rate of chromosomal rearrangements and a more muted isochore structure. However it remains to be established whether this correlation cannot be explained away simply by the increase in the overall rate of molecular evolution in the mouse lineage.

The most intriguing result generated by our analysis is the apparently coincidental increase of the CpG−based transition rate (possibly due to an increase in methylation rate of cytosine) and the decrease in the GC−dependence of the single−nucleotide transitions and transversions (possibly due to the increase in chromosomal rearrangement activities). Although we do not have an explanation for this coincidence, we would like to point out a possible causal connection: If the mammalian lineage experienced an invasion of very active transposable elements (such as L1s) at this time, the increase in methylation rate (and thus an increase in the CpG−based transition rate) could be an evolved response to control the expansion of transposable elements (Yoder, Walsh, and Bestor 1997). The increase in the rate of chromosomal rearrangements then can be due to the increase in the number of dispersed homologous sequences generated by the active transposable elements and the consequent increase in the rate of ectopic recombination.

Although the suggested explanation is highly speculative, it has the virtue of being a parsimonious, single−cause explanation for the multiple coincident changes in genomic patterns of molecular evolution in the ancestor of eutherian mammals. Further research will shed light on the validity of this model. The most robust part of this study, however, is the determination of the history of genomic patterns of molecular evolution − this history will need to be incorporated into the future accounts of the evolution of the human genome.

**Acknowledgements.** This collaboration was initiated during the workshop on "Statistical Physics and Biological Information" at the Institute for Theoretical Physics in Santa Barbara. This work was supported in part by the NSF through grant No. DMR0201103 (PA,TH) and the Stanford University OTL (DP). PA further acknowledges the support of a DFG post–doctoral fellowship, and TH a Burroughs–Wellcome innovation award in functional genomics.

**Lesefehler**

Table 1.

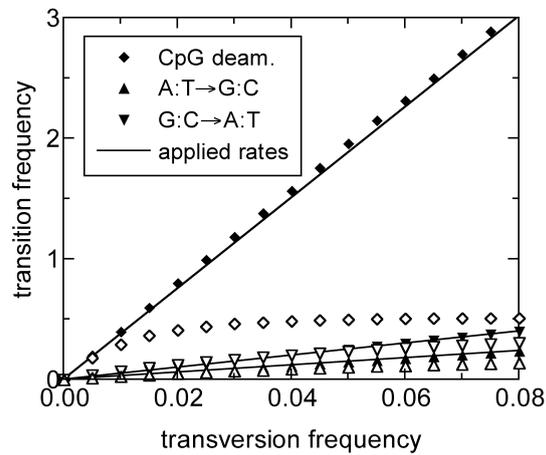

**Figure 1.** Comparison of the maximum likelihood (ML) and the direct counting (DC) method: 20,000 copies of MIR sequence (5Mbp in length) were synthetically evolved for various periods of time from an ancestor sequence. We use a stochastic model with mutation rates as given in the text.. The estimated transition frequencies are plotted against the average of the estimated transversion frequencies, for sequences evolved for different periods of time. The two single–nucleotide transition processes are shown as triangles and the CpG–based transition as diamonds. Furthermore, the ML results are distinguished by the solid symbols from the DC results (open symbols). The solid lines correspond to the actual substitution frequencies used. The ML method is able to recover the "true" substitution frequencies (the solid lines) throughout the studied regime, while the DC method quickly saturates for the fastest CpG–based transition.

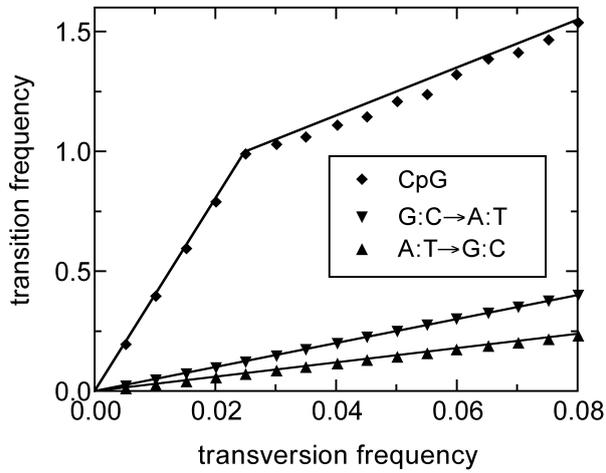

**Figure 2.** Performance test with time−dependent substitution rates: 20,000 copies of MIR consensus sequence (5Mbp in length) were aged according to a stochastic evolution model which includes a change of the CpG−based transition rate at a time $t_0 = 0.025$, similar to what is observed in Fig 4. The relative A:T→G:C rate is 3 and G:C→A:T rate is 5 as in Fig. 1. The relative CpG−based transition rate is 40 for $t < t_0$ and 8 for $t > t_0$. The data points represent the estimated substitution frequencies determined by the ML method. Note that the slopes of the data points for the 3 shown processes provide reasonable estimates of the true rates (slopes of the solid lines) throughout the time period studied.

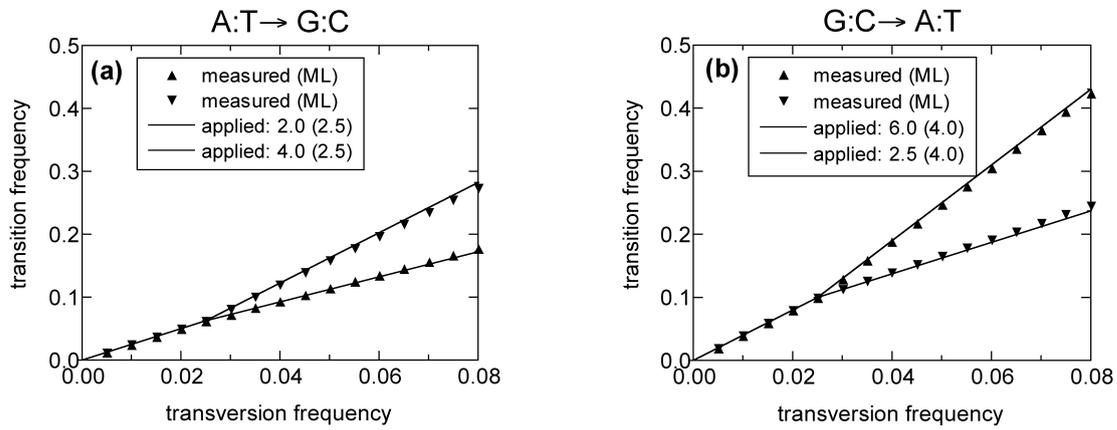

**Figure 3.** Same test as in Fig. 2, but including changes in the single−base transition rates at $t_0=0.025$, mimicking a situation found for the human genome as shown in Fig 5. The graphs show the substitution frequencies (triangles) for (a) the A:T→ G:C transition and (b) the G:C→ A:T transition as estimated by the ML analysis. Each graph presents two sets of data (the upward and downward triangles) corresponding to substitution frequencies observed in GC−poor and GC−rich regions. The transversion rates remained constant in all sets. The relative transition rates used before and after $t_0$ are indicated by the slopes of the solid lines drawn.

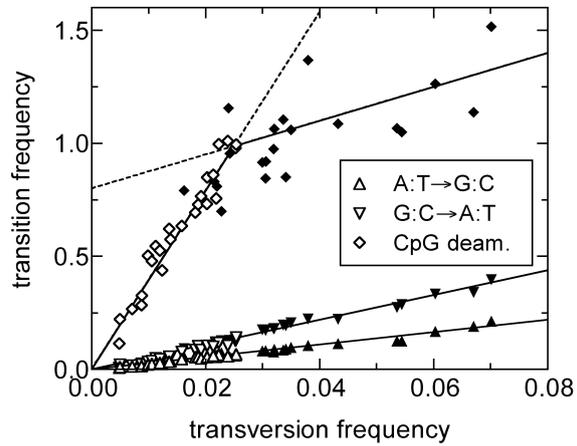

**Figure 4.** Genome−wide averages of the substitution frequencies for the 42 subfamilies of REs in human are determined using the ML−analysis. Plotted are the substitution frequencies for the two single−nucleotide transitions (triangles) and the CpG−based transition (diamonds) vs. the average transversion frequency. Each symbol represents the frequencies determined from one subfamily. Statistical error is of the size of the symbol and not shown. Filled symbols are used to represent REs that have been inserted before the mammalian radiation and are found in all mammals. The two single−nucleotide transition processes can be well fitted by a single line (R=0.99), while the CpG−based transition must instead be fitted by at least two lines joining at a time $t_0=0.025$.

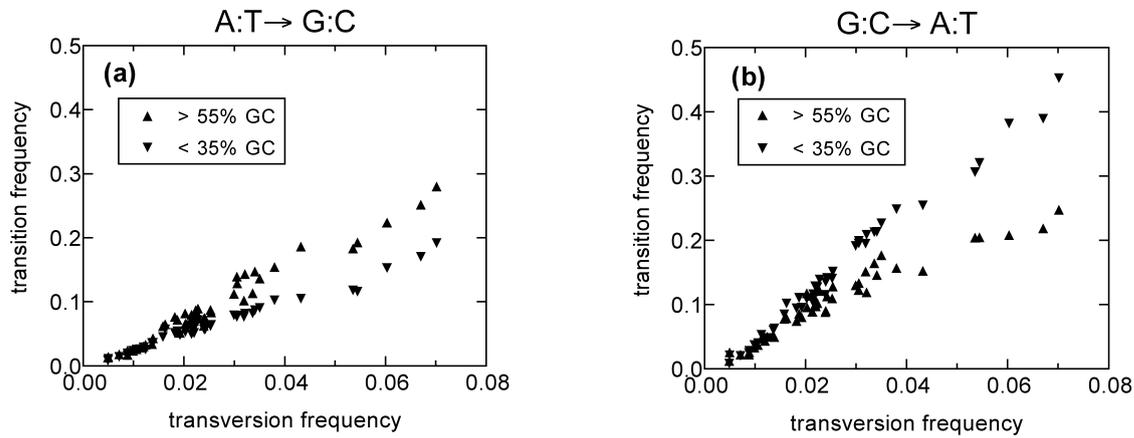

**Figure 5.** The two transition frequencies, A:T → G:C (a) and G:C → A:T (b), as determined by ML from REs found in regions with GC−contents above 55% (up−point triangles) and below 35% (down−pointing triangles). The transition frequencies show little dependence on the regional GC−content for young REs with transversion frequencies below 0.02. However for the older REs (e.g., with transversion frequencies above 0.03), the differences in the transition frequencies in GC−poor and GC−rich regions are noticeable. The GC−dependence of the transition frequencies for the older REs are *opposite* for the two transition processes.

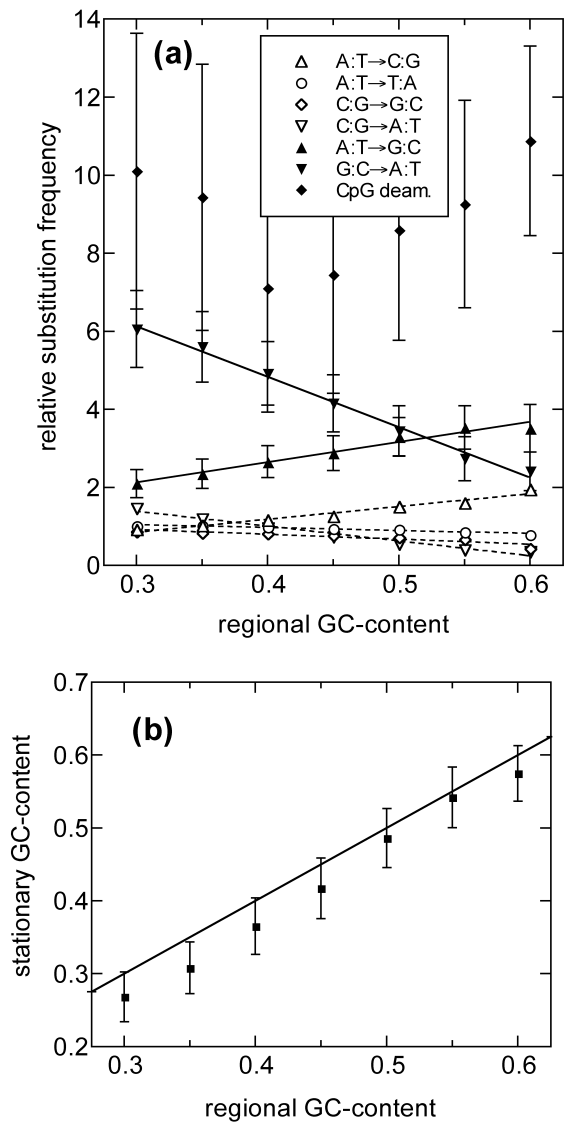

**Figure 6.** Substitution patterns prior to the "time" $t_0$: (a) substitution frequencies (relative to the average transversion frequencies) as determined from REs with different regional GC–content. These frequencies were determined by straight line fits to data such as those shown in Fig. 4, for the portion with average transversion frequencies >0.03. Up–pointing triangles represent GC–enriching substitutions and down–pointing triangles AT–enriching substitutions. Lines are drawn to guide the eyes. (b) The expected *stationary* GC–content computed for the substitution rates shown in (a) for the different regional GC–contents. The straight line indicates the equality of the stationary and regional GC–content.

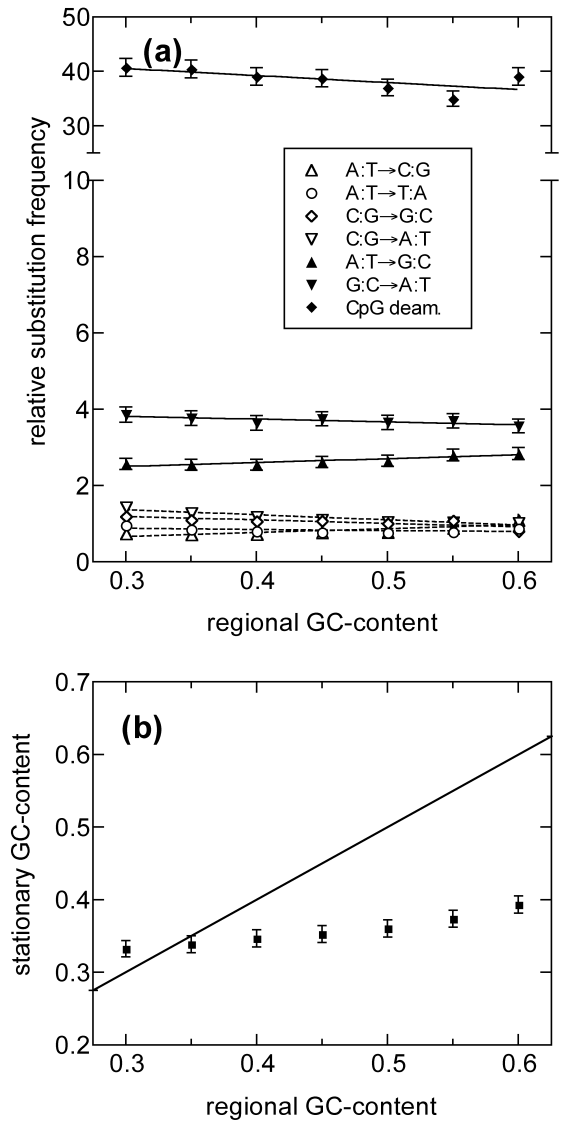

**Figure 7.** Substitution pattern after $t_0$: (a) relative substitution frequencies determined and presented in the same way as in Fig. 6(a), but using only REs with average transversion rates <0.02. Note the change in scale for the much larger CpG−based transition (diamonds). (b) The expected *stationary* GC−content computed using the GC−dependent substitution rates shown in (a). Note that it is far from the straight line (the equality of the stationary and regional GC−content), indicating that the regional GC content is far from being equilibrated by the current rates. The eventual stationary GC−content according to the plot is *35−40%*.